\documentclass[12pt,russian,american]{article}
\usepackage[T2A,T1]{fontenc}
\usepackage[latin1]{inputenc}
\usepackage{geometry}
\usepackage{graphicx}
\usepackage{setspace}
\usepackage{babel}
\begin{document}

\title{Setting of the predefined multiplier gain of a photomultiplier.}

\author{O. Smirnov}
\maketitle
\begin{abstract}
A method to set the predefined PMT gain with high precision is described.
~The method supposes minimal participation of the operator. After
the rough initial setting of the gain, the automated system adjusts
the PMT gain within a predefined precision (up to 2\%). The method
has been successfully applied at the Borexino PMT test facility at
LNGS for the gain adjustment of the 2000 photomultipliers for the
Borexino detector
\end{abstract}

\section{Introduction}

A large scale liquid scintillator detector Borexino \cite{Borex}
now under construction at the Gran Sasso underground laboratory (LNGS),
is a low energy solar neutrino detector. The scintillation from the
recoil electrons will be registered by 2200 photoelectron multiplier
tubes (PMTs) placed around a transparent inner vessel containing a
scintillating mixture. In addition 200 PMTs will be used as the external
muon veto system. 

Before installation in the detector, all PMTs are tested in the special
facility. The procedure of the testing includes preliminary gain adjustment.
In order to accelerate the high voltage adjustment an automatic system
have been developed for the PMTs test facility. Though the algorithm
of the HV adjustment was developed for the existing facility, it can
be adapted for any similar system. 

The tuning of the method has been performed during testing of 100
PMTs for the Counting Test Facility detector (the prototype of the
Borexino detector, \cite{CTF}). In August 2001 bulk testing of 2000
PMTs for the Borexino has been completed. The high efficiency of the
method permitted us the HV adjustment and complete PMT testing in
4 months. 

\section{PMT ETL9351 and its characteristics}

The Monte Carlo simulation of the Borexino detector showed that the
mean number of the photoelectrons (p.e.) registered by one PMT in
the scintillation event will be in the region $0.02-2.0$ for an event
with energy of 250-800 keV. Hence the PMTs should demonstrate a good
single electron performance. After preliminary tests, the Thorn EMI9351
with a large area photocathode (8'') has been chosen. The PMT of
this model has 12 dynodes with a total gain of $k=10^{7}$. The transit
time spread of the single p.e. response is $1-1.5$ ns. The PMT has
a good energy resolution characterized by the manufacturer by the
peak-to-valley ratio. The manufacturer (Electron Tubes Limited\c{ }
ETL) guarantees a peak-to-valley ratio of 1.5.

\begin{figure}
\caption{\label{fig: divider}High voltage divider used in the Borexino experiment.}

\centering{}\includegraphics[scale=0.75]{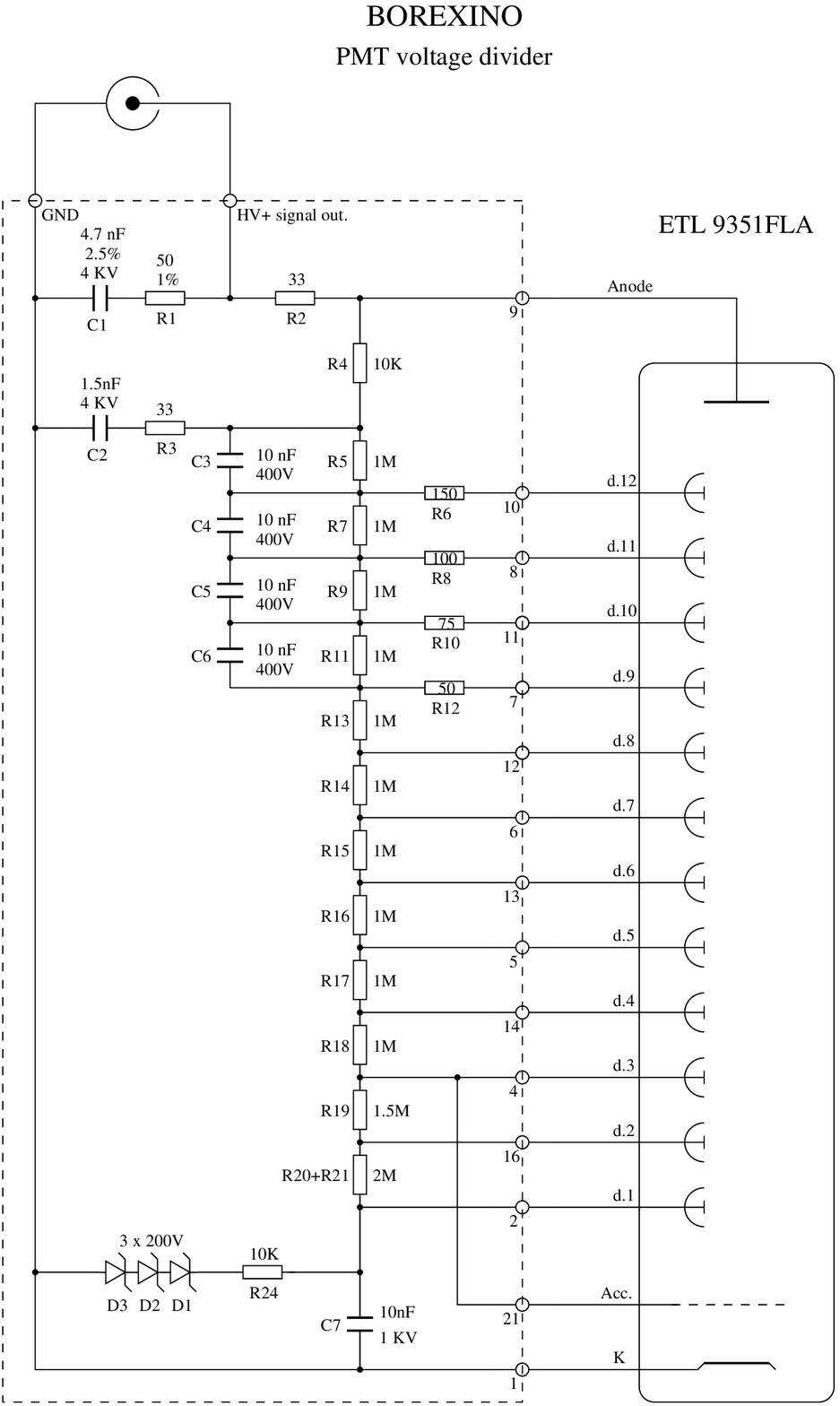}
\end{figure}

The PMTs delivered from the supplier are factory tested and the operational
high voltage (HV) is specified by the manufacturer. Selective measurements
of the PMTs at the factory specified voltage showed a high variance
of the gain $k=(0.86\pm0.25)\times10^{7}$. The high voltage divider
used by ETL and the one used in Borexino are different. For these
reasons the voltage should be remeasured. The Borexino PMT divider
is shown in fig.\ref{fig: divider}. For proper signal termination
the 50$\Omega$ resistor R1 is included in the scheme. The resistor
decreases signal amplitude by a factor of 2. The signal attenuation
is compensated by the higher operating voltage. Practically the photoelectron
multiplier is operating at the $k=2\times10^{7}$ gain.

\section{PMT test facility at LNGS}

In the frame of the Borexino programme the special PMT test facility
was prepared at the LNGS. The test facility is placed in two adjacent
rooms. In one room the electronics is mounted, and the other is a
dark room with 4 tables designed to hold up to 64 PMTs. The dark room
is equipped with an Earth's magnetic field compensation system on
the base of rectangular coils with an electric current (\cite{5}).
The non-uniformity of the compensated field in the plane of the tables
is no more than $10\%$. The tables are separated from each other
by black shrouds, which screen the light reflected from the PMTs photocathode.

\begin{figure*}
\caption{\label{fig: blockscheme}One channel of the electronics used in the
measurements}

\centering{}\includegraphics[scale=0.75,angle=90]{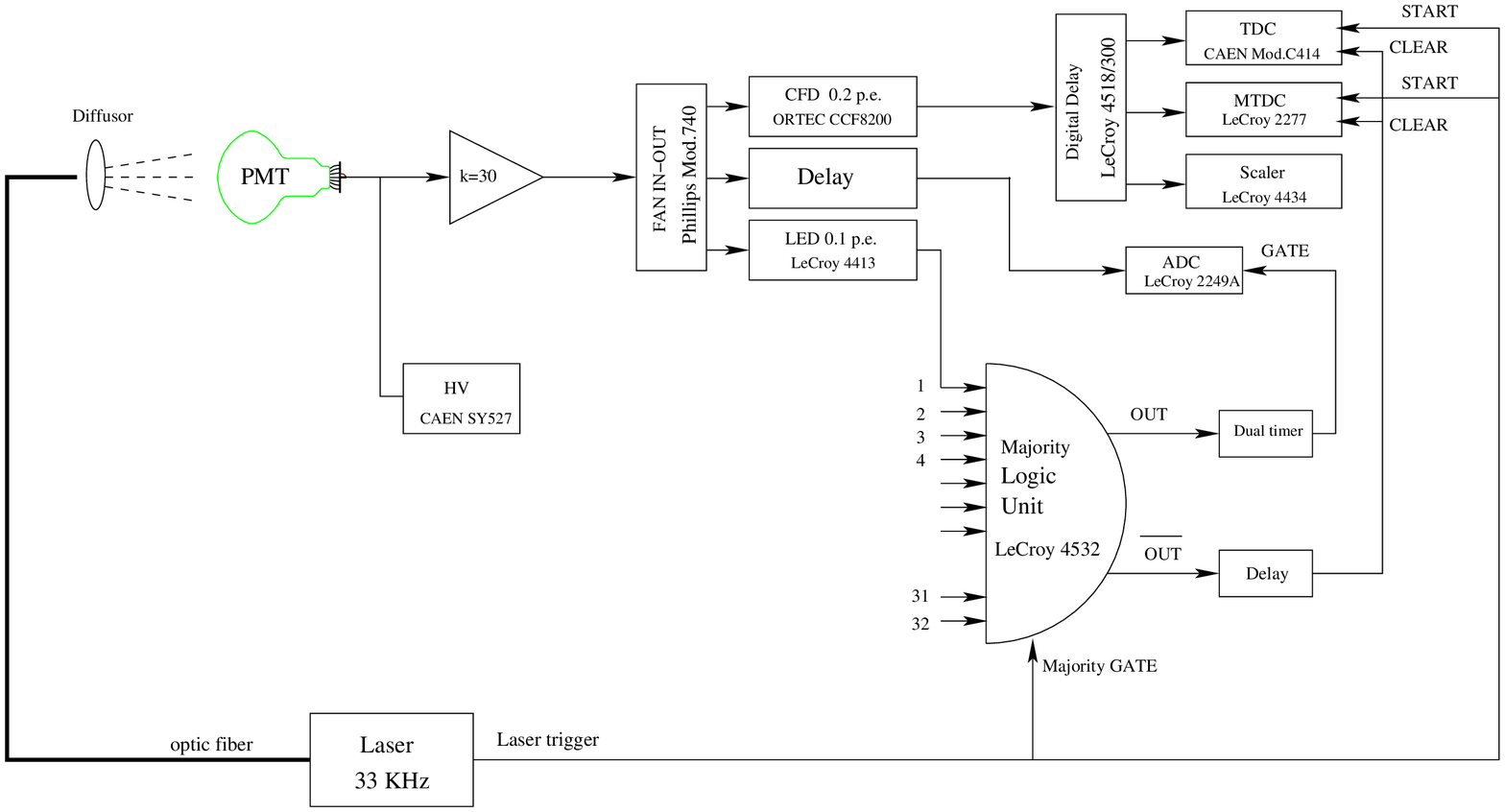}
\end{figure*}

One channel of electronics (out of the total 32) of the test facility
is presented in fig.\ref{fig: blockscheme}. The system uses the modular
CAMAC standard electronics and is connected to a personal computer
by the CAEN C111 interface. The PMT characteristics are defined during
a 5 hour run. The stability of the parameters is defined every 12-24
hours during longer runs.

The PMTs are illuminated by low intensity light pulses from a laser.
A picosecond Hamamatsu pulse laser was used in the tests. The model
used has a peak power of $0.39\:mW$, the pulse width is $27.3\:ps$,
and the laser wavelength is $415\:nm$, which is close to the maximum
quantum efficiency of Thorn EMI 9351. The light pulse from the laser
is delivered by a 6 meter long optical fibers into the dark-room.
Each of the 4 fibers is supplied with a diffuser in order to provide
a more uniform illumination of the tables.

The ADC gate and TDC ``start'' signals are generated using the laser
internal trigger\c{ } which has negligible time jitter ($<100\:ps$)
with respect to the light pulse. The ``stop'' signal for the TDC
is formed by the constant fraction discriminator (CFD) with the threshold
set at the 0.2 p.e. level.

\begin{figure*}
\caption{\label{fig: example}The characteristics of one of the PMTs under
test: charge spectrum (upper plot) and\"{ }afterpulses (lower plot).
The PMT charge response is measured in ADC channels (1024 ADC channels
corresponds to 256 pC). The peak at $\sim30$ $\mu$s in the lower
plot corresponds to the laser frequency.}

\begin{centering}
\includegraphics[scale=0.5]{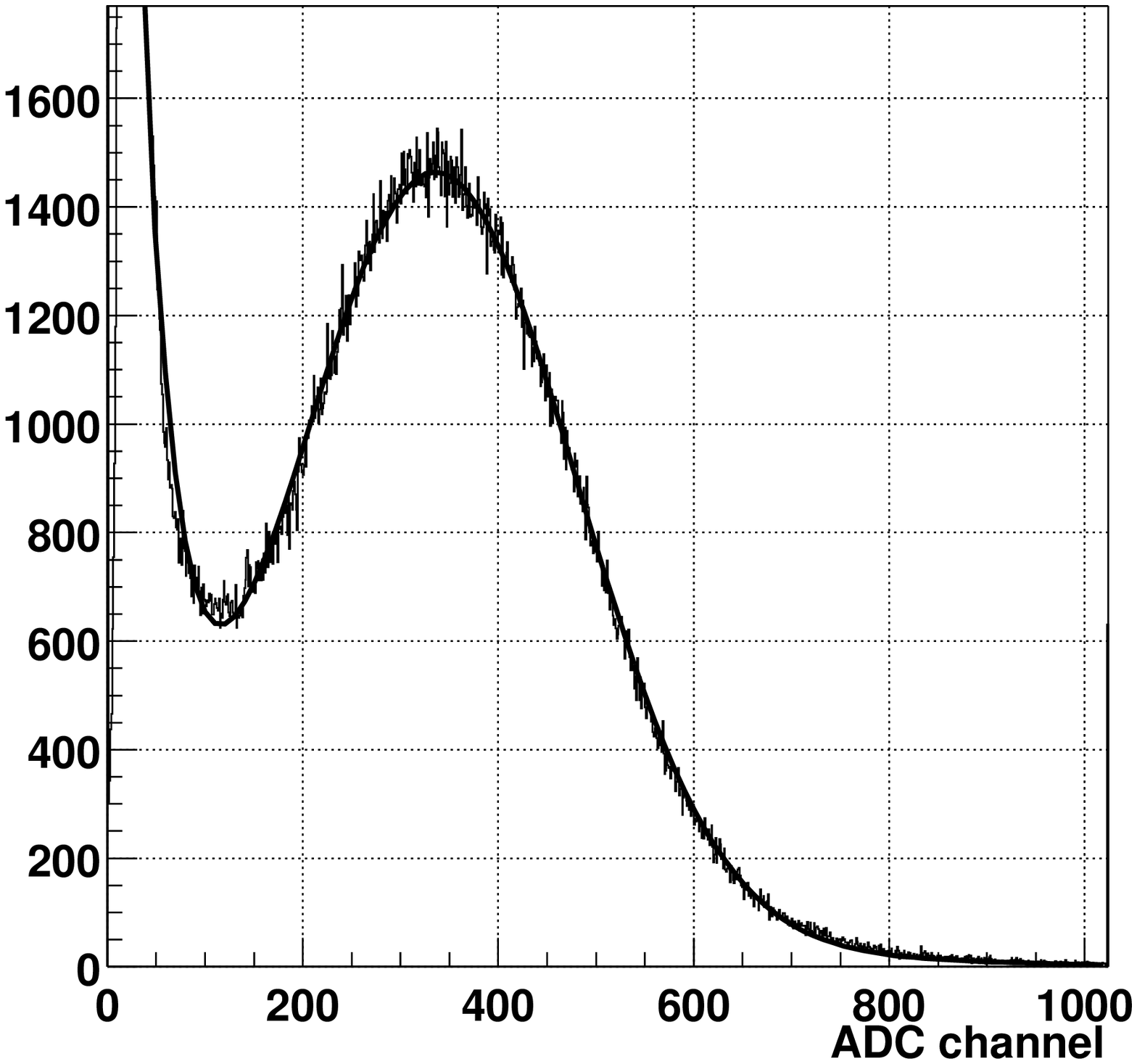}
\par\end{centering}
\centering{}\includegraphics[width=0.8\textwidth,height=0.25\textwidth]{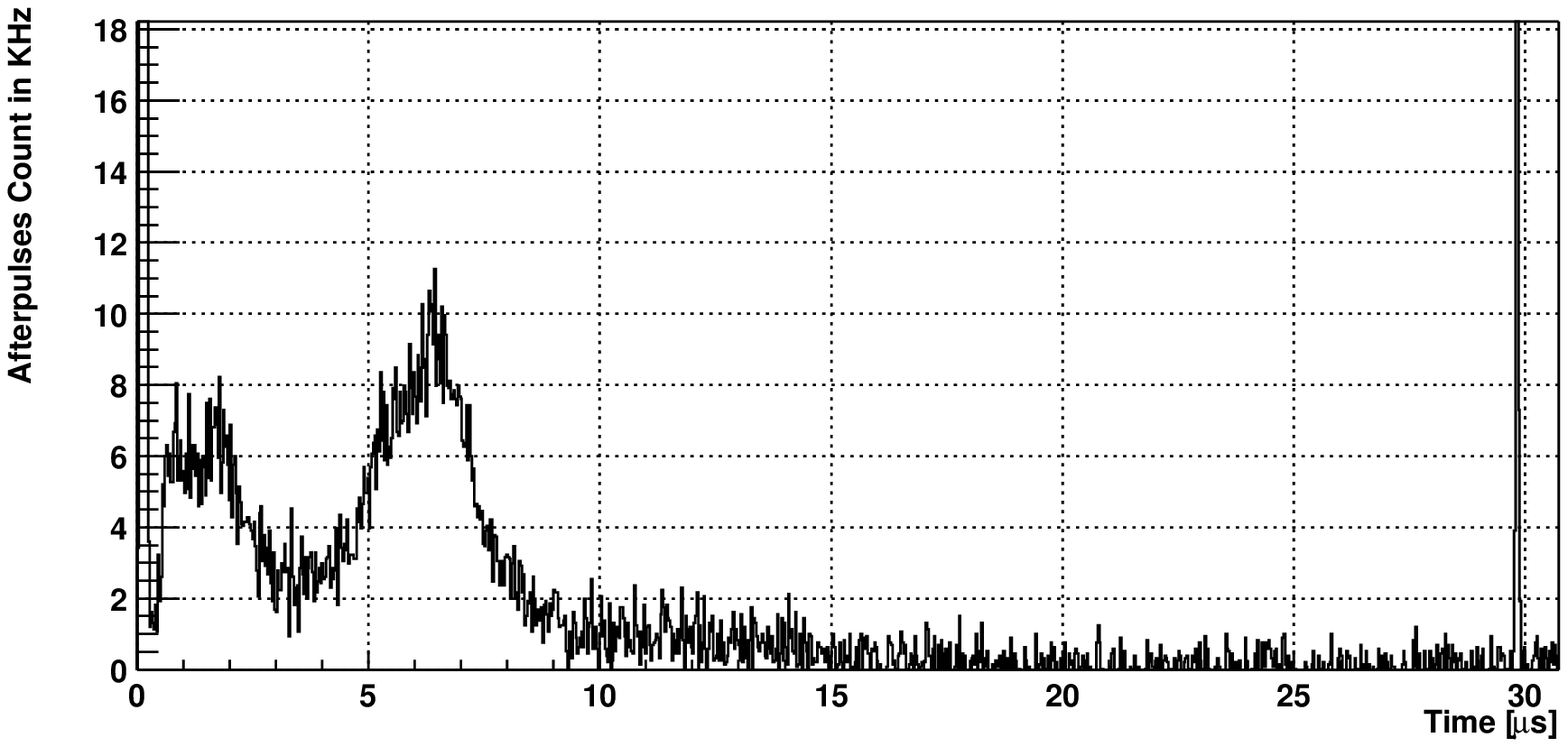}
\end{figure*}

The 32-inputs majority logic unit is able to memorize the pattern
of the hit channels. This information significantly increases the
data processing rate. The reading of the electronics is activated
when the majority LAM signal is on (a LAM is produced if one of the
signals on the inputs is inside the external GATE on the majority
logic unit). Otherwise, a hardware clear is forced using the majority
$\overline{OUT}$ signal. Every pulse of the laser is followed by
an internal trigger. The trigger is used as the majority external
gate. An example of the data acquired during a routine PMT test is
presented in fig.\ref{fig: example}. It should be pointed out that
the charge histogram in fig.\ref{fig: example} is acquired together
with the ADC pedestal, i.e. without a hardware threshold. This was
realized by connecting the last (32-th) majority input to the external
gate signal. In this case the system is triggered at the first trigger
from the laser that occurs when the electronics is not busy with the
previous data transfer. During the HV tuning, the ``cut'' charge
histograms are acquired with a hardware threshold of about $5\%-10\%$
of the ``typical'' Single Photoelectron Response (SER) mean value\footnote{signals at the majority inputs are formed by the Leading Edge Discriminator
(LED) with a threshold set to this value. A Constant Fraction Discriminator
(CFD) with a higher threshold (about $20\%$ of the ``typical''
SER mean value) is used for the timing measurements.}. 

The afterpulses are registered by the multihit TDC (MTDC) which is
able to memorize in the internal register up to 16 hits inside a 32
$\mu s$ window.

The high precision calibration of each electronics channel had been
performed before the measurements. Here calibration means the ADC
response to a signal corresponding to 1$\:$p.e.\footnote{multiplied by a factor of $10^{7}$ by the electron multiplier and
giving $1.6\:$pC charge} on the system input with the ADC pedestal subtracted (the PMT in
this measurement is substituted by a precision charge generator LeCroy
1976). The position of the ADC pedestals are defined and checked during
the run.

\section{Electron multiplier gain measurements}

For the HV tuning it is necessary to provide a robust method of the
gain measurement. The method of a multiplier gain measurement with
multichannel analyzer in the self- triggering mode has a precision
worse than $10\%$ \cite{Wright}. The fundamental limit comes from
the inability of a multichannel analyzer used in the measurements
to access all the contributions from the low amplitude region. The
systematic uncertainty of the methods using the SER modeling (see
i.e.\cite{Chirikov-Zorin}) are difficult to estimate, especially
during bulk testing. Another disadvantage of these methods is the
high statistics necessary to estimate the SER parameters. 

A method of photomultiplier calibration with a high precision of up
to a few percent has been discussed in the article \cite{Filters}.
The method is based on precision measurements of the PMT charge response
to low intensity light pulses from a laser. It has been concluded
that the precision of the method is limited only by the systematic
errors in the discrimination of the small amplitude pulses from the
electronics noise. On the basis of our experience with the precise
PMT calibration \cite{Filters}, a fast procedure of PMT voltage tuning
has been realized for the Borexino PMT test facility at the Gran Sasso
laboratory.

\begin{figure*}
\caption{\label{fig:HV}Dependence of the PMT gain on the HV applied. The HV
is measured in Volts, the PMT gain is measured in units of $10^{7}$.
The straight lines on the plot are calculated using formula$\:$(\ref{formula:dk})
with n=3.5. }

\begin{centering}
a)\includegraphics[scale=0.4]{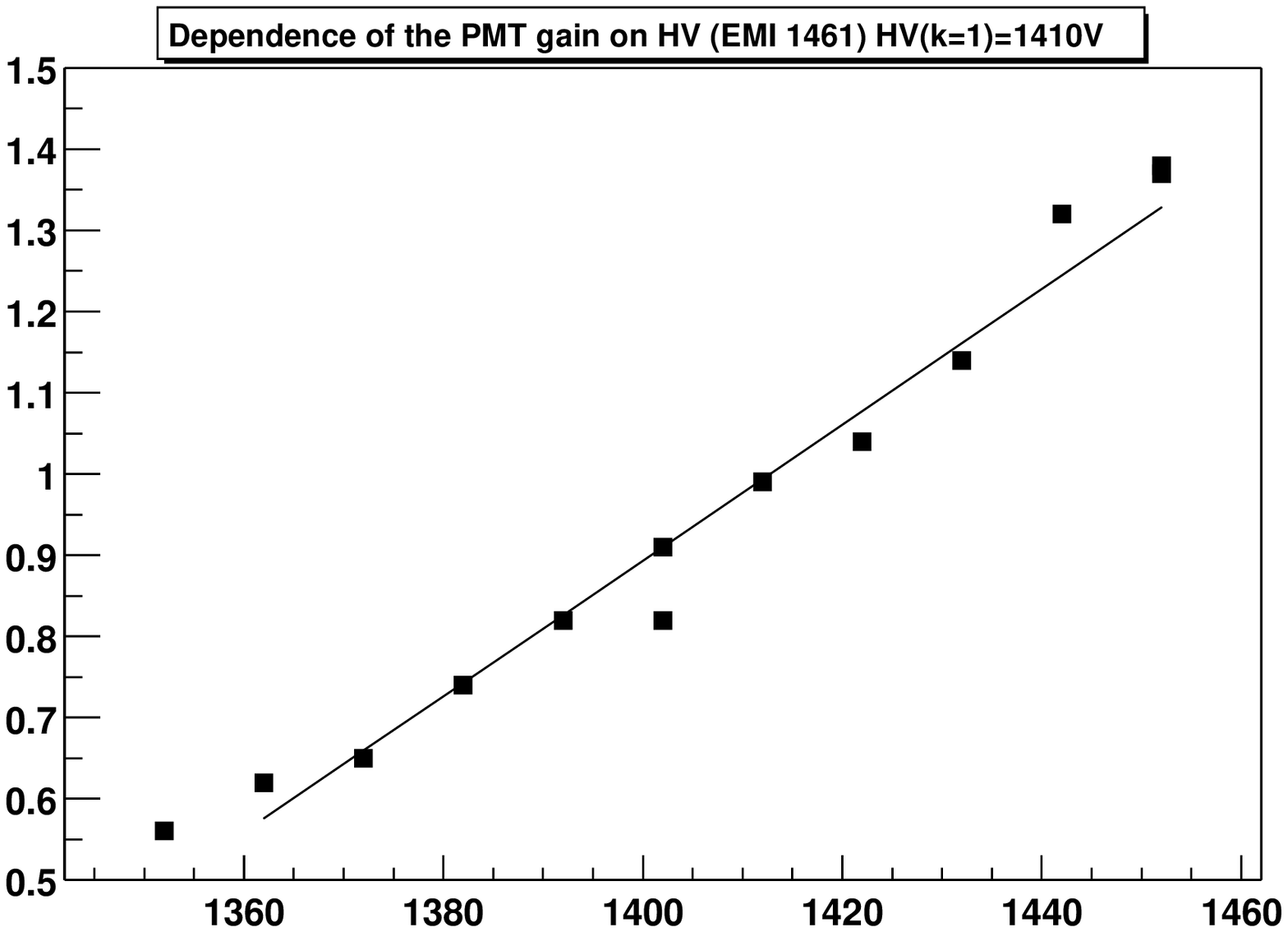}b)\includegraphics[scale=0.4]{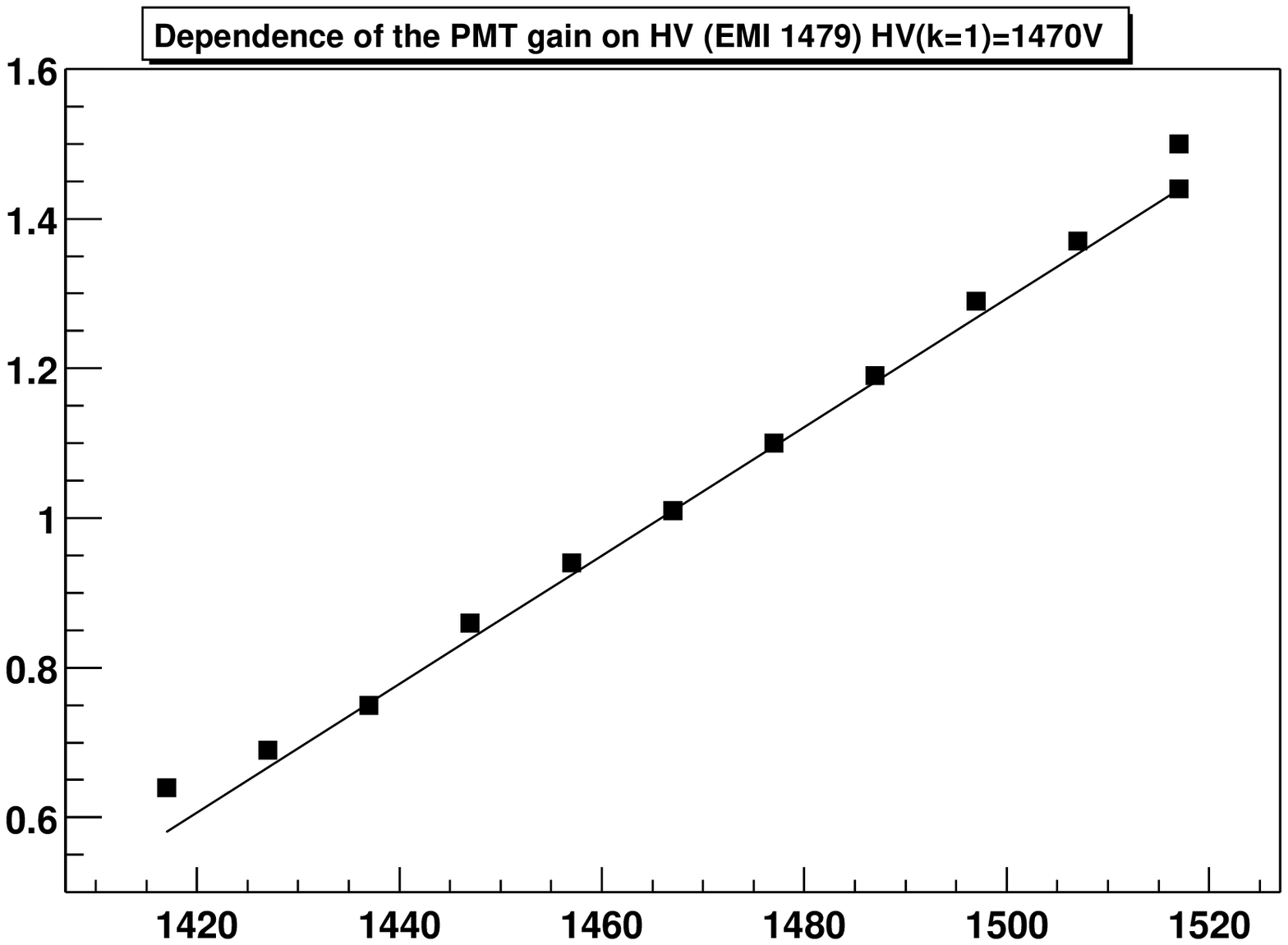}
\par\end{centering}
\centering{}c)\includegraphics[scale=0.4]{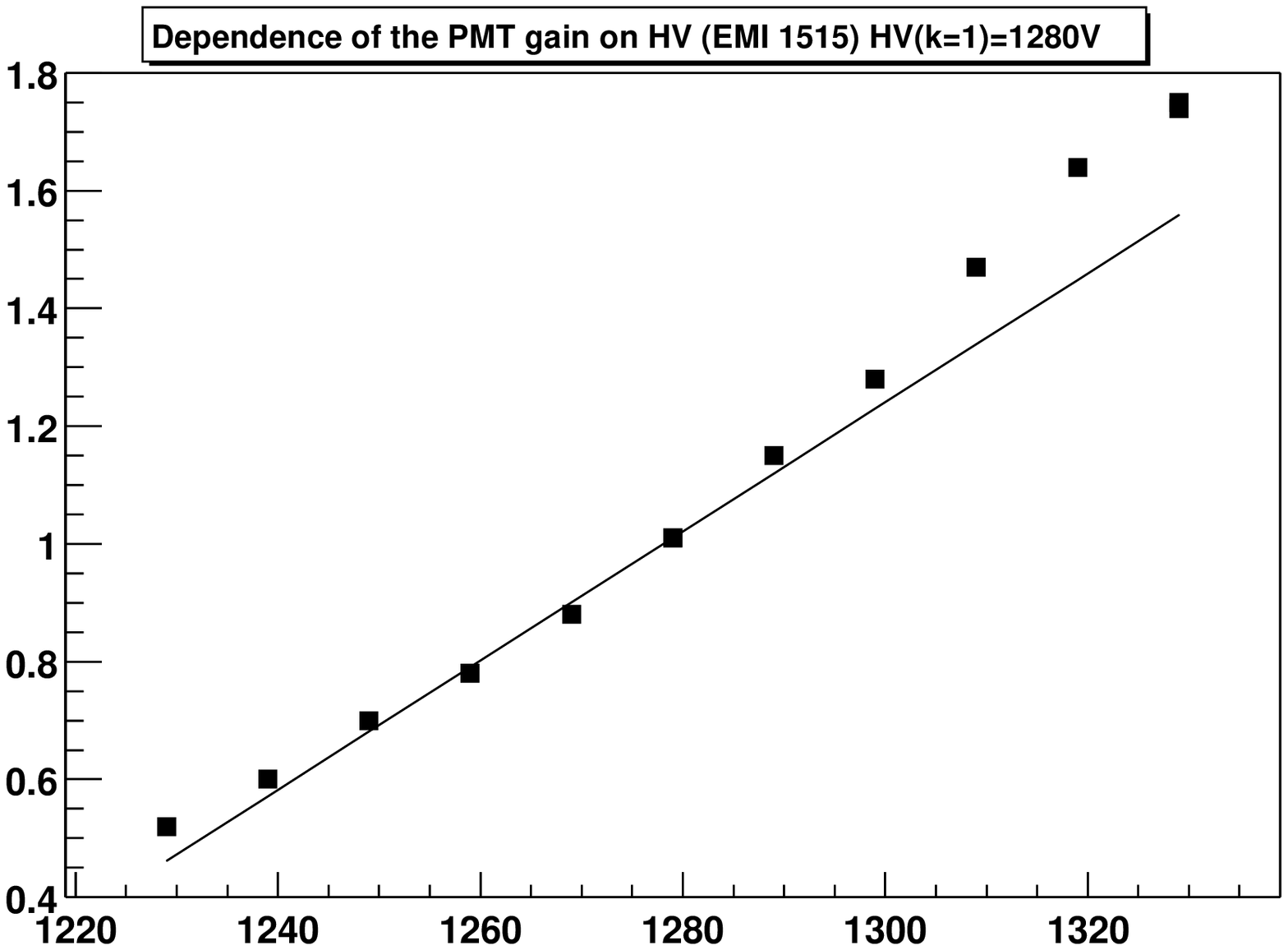}d)\includegraphics[scale=0.4]{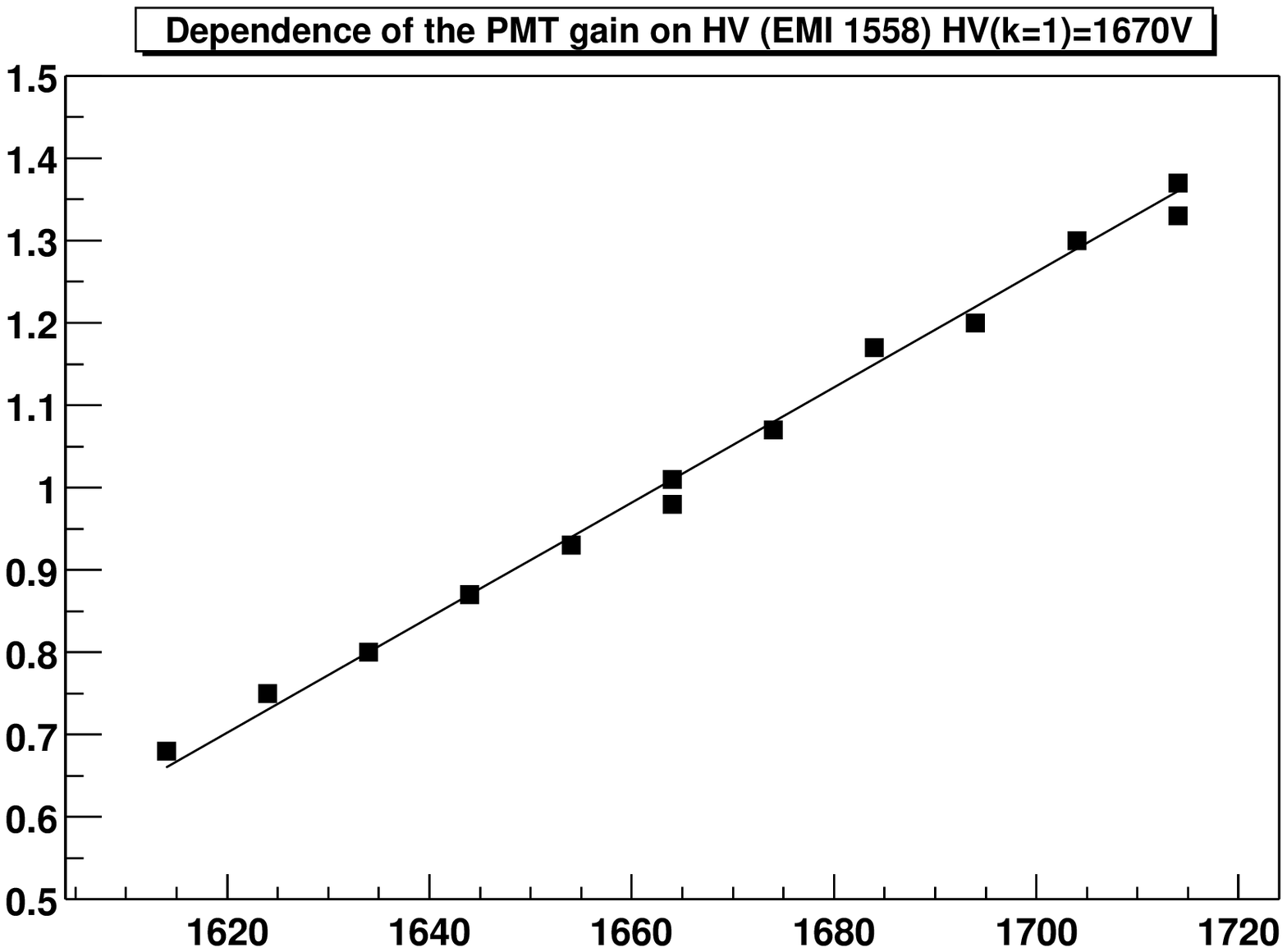}
\end{figure*}

The goal of the fine tuning is to find the HV value that will provide
$k=10^{7}\:$electron gain factor for each PMT. The mean value of
the charge SER$\:\:q_{1}$ is determined, and the HV is adjusted so
that $q_{1}$ agrees with a calibration value $c_{1}=1.6\:pC$ to
a predefined precision. Because of the hardware cut in the charge
data, the following correction should be performed in order to obtain
the $q_{1}$value from the cut distribution (see Appendix \ref{sec: appendix}):

\begin{equation}
q_{1}=q_{m}\frac{1-P(0)\cdot(1+\mu\cdot p_{t}\,)}{\mu}(1-P(0)p_{t}\frac{thr}{2})^{-1},\label{eq:q1corrected}
\end{equation}

here $q_{m}$ is the mean value of the cut distribution (a software
cut of $15\%$ of $c_{1}$ is used in order to avoid the effects of
the SER shape distortion near the hardware threshold);

\begin{figure*}
\caption{\label{Fig:Mu}The mean number of the photoelectrons registered for
one laser shot.}

\selectlanguage{russian}%
\centering{}\includegraphics[scale=0.5]{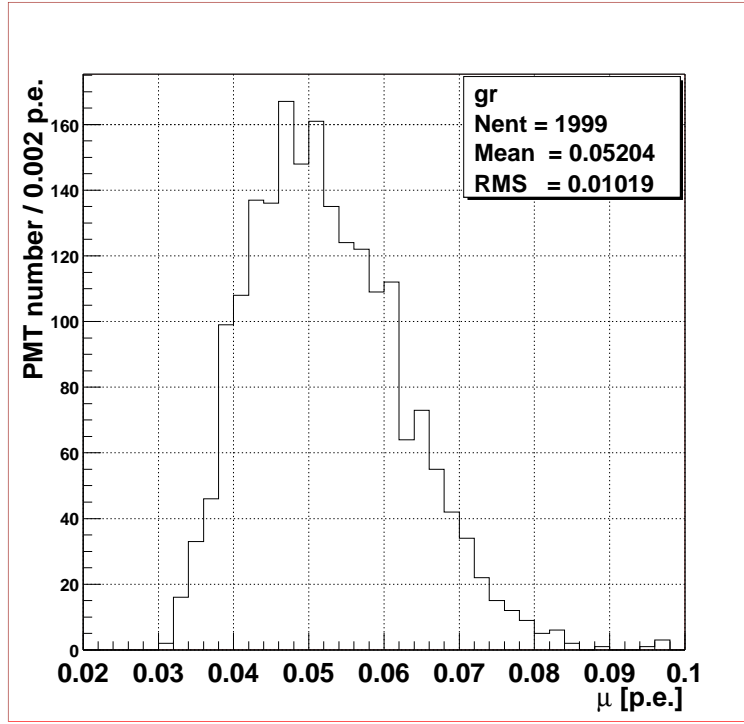}\selectlanguage{american}%
\end{figure*}

$\mu$ is the mean p.e. number registered for one laser pulse (see
fig.\ref{Fig:Mu});

\begin{figure*}
\caption{\label{Fig:pt} $p_{t}$ parameter defined during the tests.}

\centering{}\includegraphics[scale=0.4]{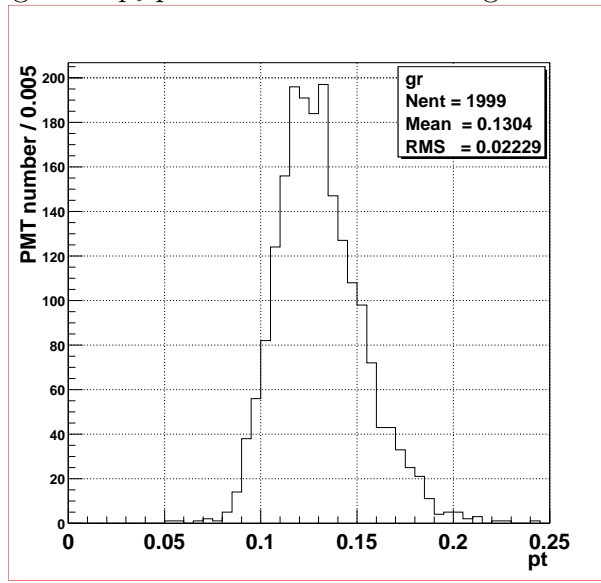}
\end{figure*}

$p_{t}$ is the part of the charge SER under the threshold. For the
15\% software threshold, the value $p_{t}\cong0.11$ was used, defined
during the tests of 100 PMTs of CTF programme (see fig.\ref{Fig:pt}); 

$thr$ is the threshold level measured in the $c_{1}$ units (0.15
in the our case).


For small $\mu$, equation (\ref{eq:q1corrected}) can be significantly
simplified:
\begin{equation}
q_{1}=q_{m}\frac{1-\frac{\mu}{2}-p_{t}}{1-p_{t}\frac{thr}{2}}.\label{eq:simplified}
\end{equation}

The mean p.e. number is defined during the test by estimating the
probability of two sequential non-zero signals on PMT. Assuming a
Poisson distribution of the light detection process, one can write:

\begin{equation}
\mu=-\ln(1-\frac{N_{ev}}{N_{Triggers}})\simeq\frac{N_{ev}}{N_{Triggers}},\label{Eq:mu}
\end{equation}

where $N_{Trigger}$ is the the full number of the system triggers
and $N_{ev}$ is the number of events that are followed by the non-zero
pulse (the first signal in a two pulses sequence is triggering the
system and thus is always present). In practice we take as $N_{Trigger}$
the number of the events in the charge histogram (i.e. with a $5\%-10\%$
hardware cut), and as $N_{ev}$ we take the number of events after
a 30$\,$$\mu s$ delay\footnote{30$\,\mu s$ corresponds to the laser repetition rate of 33 kHz. The
laser repetition rate is fixed in the measurements at this value.} estimated from the Multihit TDC histogram (see fig.\ref{fig: example}).
For these events the hardware cut on the CFD is about $20\%$ of the
SER. The precision of the $\mu$ estimation using these $N_{Triggers}$
and $N_{ev}$ values is approximately $10\%$, which is good enough
for our purpose.

\section{The dependence of the multiplier gain on the HV}

\begin{figure*}
\caption{\label{Fig:HV}Results of the HV tuning. }

\centering{}\includegraphics[scale=0.5]{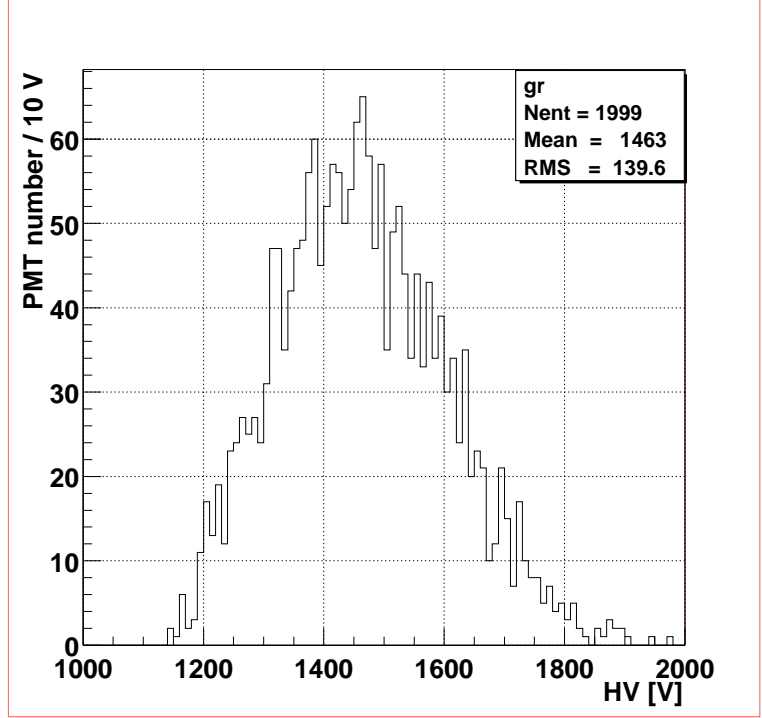}
\end{figure*}

The PMT divider (fig.\ref{fig: divider}) provides a fixed voltage
difference between the photocathode and the first dynode ($U_{D1}=600\:$V).
The remaining potential is distributed between 11 dynodes: $U=U_{D2}+U_{D3}+U_{D4}+...+U_{D12}=13.5U_{0},$
with the last 9 voltage steps being equal, $U_{0}=\frac{U-600V}{13.5}$,
while $U_{D2}=2U_{0}$ and $U_{D3}=1.5U_{0}$ respectively. For the
typical PMT the HV value is in the 1200-1700$\:$V range (see fig.\ref{Fig:HV}),
i.e. $U_{0}<80$ V. The Be-Cu dynodes of the EMI$\:$9351 have a gain
that changes linearly with the applied voltage up to 200-250 V (see
fig.9 of \cite{Cat}). If the electron multiplication on the first
dynode is $g_{1}$ and amplification of the dynode at the $U_{0}$
voltage is $g=k_{d}\cdot U$, then the total gain of the 12-dynodes
system is: 
\begin{equation}
k=3\cdot g_{1}\cdot(k_{d}\cdot\frac{U-600}{13.5})^{11-n}g_{n},\label{Formula:k}
\end{equation}
where n is introduced in order to take into account the influence
of the spatial charge on the last dynodes, where the dynode gain is
practically independent of the applied high voltage. The total gain
on the last dynodes is assumed to be $g_{n}$. The relative variation
of the gain versus the variation of the applied voltage is:
\begin{equation}
\frac{dk}{k}=(11-n)\cdot\frac{dU}{U-600}.\label{formula:dk}
\end{equation}

This equation gives an exponential law for the gain factor as a function
of applied voltage. 

In order to check the dependence of the PMT gain on the applied high
voltage, a special set of measurements has been performed for 4 PMTs
with different operational voltages. The results are presented in
fig.\ref{fig:HV}. It was found that $\frac{dk}{k}$ is better described
by a straight line calculated with formula$\:$(\ref{formula:dk})
with a constant factor $n\simeq3.5$.

\section{HV tuning procedure}

The HV tuning procedure starts with the HV set to the minimum value
of 1200 V. Then the HV for the each PMT is increased in order to achieve
a dark count rate of $\approx1000\:s^{-1}$. During all the operations
the PMT current is checked, and if it is too big or unstable the PMT
is switched off. 

After the initial adjustment of the HV, a short data acquisition cycle
is performed. PMT charge histograms are inspected visually one by
one, and further HV adjustment is applied as necessary (i.e. if the
position of the charge SER mean value is too low or too high), aiming
to achieve $k=10^{7}\pm80\%$. Then the fine HV tuning is performed
in a special mode of the data acquisition.

For the following discussion the relation between the mean value of
the SER and its r.m.s. is important. The relative variance $v_{1}=(\frac{\sigma_{q_{1}}}{q_{1}})$$^{2}$
of the SER charge spectrum was estimated during the tests (see$\,$\cite{Database}
for details). The maximum value of this parameter for the PMTs under
test is about 0.4; hence we used this value for the $\sigma_{q_{1}}$
estimation ($\sigma_{q_{1}}=\sqrt{v_{1}}q_{1}$). For the spectrum
of $N_{Triggers}$ events the statistical precision of the mean charge
estimation is $\sigma_{q}=\frac{\sigma_{q_{1}}}{\sqrt{N_{Triggers}}}$
(not taking into account that the precision gets worse because of
the uncertainties introduced by the spectrum cut).

\begin{figure*}
\caption{\label{fig: hv_chart}The flow chart of the high voltage tuning}

\centering{}\includegraphics[scale=0.8]{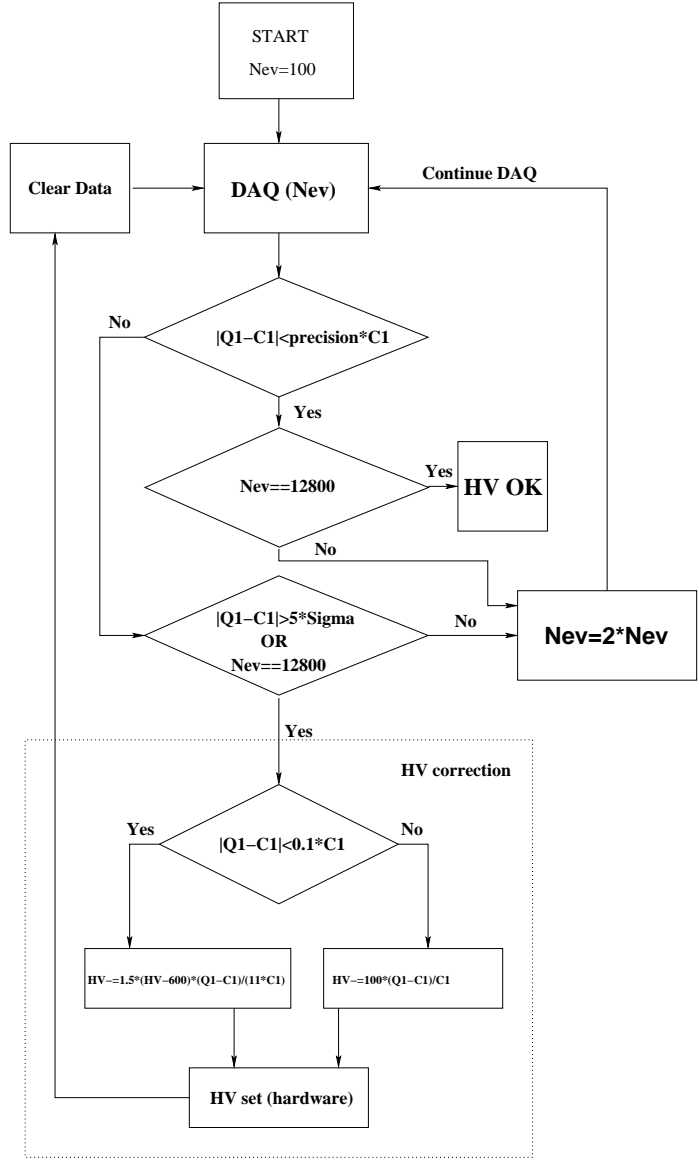}
\end{figure*}

The fine tuning starts with the acquisition of $N_{Triggers}=100$
events, then the $q_{1}$ is estimated ($\mu$ is not estimated in
the first stages and is assumed to be 0.05$\;$p.e.). If the $q_{1}$
value is inside the $c_{1}\pm5\sigma_{q}$ interval, the maximum number
of the events is doubled and another $N_{Triggers}$ events are acquired
in order to increase the statistics. If $\mid q_{1}-c_{1}\mid>5\sigma_{q}$
the HV is adjusted, the data are cleared, and the data acquisition
starts again for the $N_{Triggers}$ events. We start with only 100
events, so that the HV is adjusted very rapidly if $q_{1}$ is far
from $c_{1}$\footnote{nevertheless, 100 events statistics provides about $20\%$ precision
of the calibration at this stage ($5\cdot\frac{0.4}{\sqrt{100}}$)}. The logic of the fine HV tuning is presented in fig.\ref{fig: hv_chart}.
The interval $5\cdot\sigma_{q}$ is chosen in order to take into account
the possible systematic errors.

The maximum (preset) number of the events is 12800, which provides
a $1\%\;$statistical precision ($\frac{\sqrt{v_{1}}}{\sqrt{12800}}<0.01$).
When the event number achieves the maximum value (i.e. 12800), the
condition $\mid q_{1}-c_{1}\mid<p_{HV}\cdot c_{1}$ is checked, where
$p_{HV}$ is the predefined precision of the HV adjustment (typically
$2\%).$ If the condition is true, the acquisition in the channel
is stopped and discriminator output is disabled, in such a way increasing
the data acquisition rate for the remaining channels.

The HV correction is calculated from the following considerations.
If the deviation is big ($>10\%)$ the correction is set to a fraction
of the maximum deviation of 100$\:$V in proportion to the deviation
from the calibration value $\frac{q_{1}-c_{1}}{c_{1}}$. The low enough
value of 100$\,$V has been used, so that any possible overvoltage
is avoided. For small deviations from the calibration ($<10\%)$ the
correction can be calculated more precisely from (\ref{formula:dk}),
namely: 

\begin{equation}
\Delta U=-\frac{3}{2}\frac{U-600}{11}\frac{q_{1}-c_{1}}{c_{1}}.\label{Eq:Delta_HV_correct}
\end{equation}

\section{Results}

We find the algorithm is sufficiently fast; for 30$\:$PMTs at 1$\:$kHz
acquisition rate the HV is adjusted in 15-20 minutes with $2\%$ statistical
precision.

\begin{figure}
\caption{\label{fig: stat}Results of the 2000 PMTs HV tuning. Multiplier gain
is in units of k=$10^{7}$.}

\centering{}\includegraphics[scale=0.4]{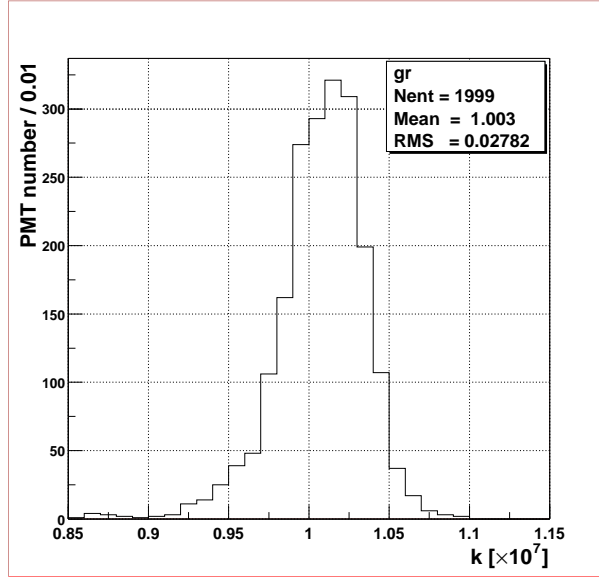}
\end{figure}

The results of the HV tuning with the precision set to $2\%\:$$\:$are
presented in fig.\ref{fig: stat}. These results were obtained during
the high precision tests after the HV adjustment. The mean value of
k agrees with the expected $k=1.0\times10^{7}$.

The systematic error of the method is connected mainly with the substitution
of the parameter $p_{t}$ by its mean value $p_{t}=0.11$, the error
caused by the precision of $\mu$ estimation is negligible. The variance
of the $p_{t}$ parameter is $\sigma_{p_{t}}=0.018$ (see fig.\ref{Fig:pt}).
As follows from the formula (\ref{eq:simplified}) the contribution
of the systematic error is equal to $\sigma_{p_{t}}$. Summing quadratically
the systematic (0.018) and the statistical (0.02) errors one will
obtain the full error of $\sigma\simeq0.027$. The variance of the
gain distribution (0.028) agrees with the calculated value. 

\section{Conclusions}

The described method is not based on any specific single photoelectron
spectrum model and hence can be used for the gain adjustment of the
PMT with an arbitrary single electron response. Other advantages of
the method are the simplicity of calculations, the predictable precision
and very high speed of the algorithm. The method can be adapted for
use with any type of PMT designed to operate in single electron regime.

The results of the high voltage tuning for the 2000 PMTs of the Borexino
experiment confirmed that the method is robust and very effective.

\section{Acknowledgements}

This job would have been impossible without the support from the INFN
sez. di Milano. I would like to thank Prof. G.Bellini and Dr. G.Ranucci
who organized my stay at the LNGS laboratory. I am grateful to Richard
Ford for the critical reading of the manuscript and useful discussions.

I would also like to thank my colleagues from the Borexino collaboration.

\appendix

\section{Appendix:\label{sec: appendix} Correction of the calibration for
the cut spectrum}

Let us calculate the mean of the PMT charge spectrum cut at a certain
level (i.e. the pedestal and the small amplitude pulses cut). The
mean registered charge in this case can be defined as: 
\begin{equation}
q_{m}=\int_{q_{th}}^{\infty}\,\sum_{N=0}^{\infty}P(N)f_{N}(q)q\,dq=\sum_{N=0}^{\infty}\int_{q_{th}}^{\infty}f_{N}\!(q)q\,dq\equiv\sum_{N=0}^{\infty}P(N)\overline{q_{N}\!(q_{th}),}\label{eq:qmean}
\end{equation}
 here $f_{N}(q)$ is the PMT response to the N photoelectrons (p.e.).
We will assume that the probability distribution for N$\:$p.e. registered
by the PMT is Poissonian. In order to take into account the fact that
a charge less than the threshold will not be registered, the Poisson
probabilities should be renormalized (more precisely, the conditional
probabilities should be calculated):
\[
P(N)\rightarrow\frac{P(N)}{1-P(0)-p_{t}\cdot P(1)}.
\]
Part of the signals under the threshold is interpreted as a no response
signal i.e. $P(0)\rightarrow P(0)+p_{t}P(1)$ and $P(1)\rightarrow P(1)(1-p_{t})$.
We assume here that the part of the PMT response under the threshold
is negligible for 2 and more p.e. registered, and all the response
to 0 p.e. remains under the threshold. The part of N=1 response under
the threshold is $p_{t}$ by definition. In this case $\overline{q_{N}(q_{th})}=N\cdot q_{1}$
for $N\geq2$. The response to N=0 p.e. has a mean value $\overline{q_{0}(q_{th})}=0$
. In order to obtain the mean value of the N=1 response we will use
the following approximation for the N=1 response (with a rectangular
part under the threshold):
\[
f_{1}(x)=\left\{ \begin{array}{cc}
0 & q<0\\
\frac{p_{t}}{q_{th}} & 0<q<q_{th}\\
Ser(q) & q>q_{th}
\end{array}\right.
\]
then $q_{1}=(1-p_{t})\overline{q_{1}(q_{th})}+p_{t}\cdot\frac{q_{th}}{2}$
and we can rewrite$\:($\ref{eq:qmean}) as

\[
q_{m}=\frac{1}{1-P(0)-p_{t}\,P(1)}\left[P(1)(1-p_{t})\overline{q_{1}(q_{th})}+\sum_{N=2}^{\infty}P(N)Nq_{1}\right].
\]
Noting that 
\[
\sum_{N=2}^{\infty}P(N)N=\sum_{N=0}^{\infty}P(N)N-P(1)
\]
and 
\[
\sum_{N=0}^{\infty}P(N)N\equiv\mu
\]
 we can finally obtain
\[
q_{m}=\frac{\mu\cdot q_{1}}{1-P(0)-p_{t}\,\mu\,P(0)}\left[1-P(0)p_{t}\frac{q_{th}}{2q_{1}}\right].
\]


\begin{thebibliography}{1}
\bibitem{Borex}Arpesella C. et al., ``Borexino at Gran Sasso - Proposal
for a real time detector for low energy solar neutrino.'' Volume
1. Edited by G.Bellini,M.Campanella,D.Guigni. Dept. of Physics of
the University of Milano. August 1991. 

\bibitem{CTF}Alimonti G. et al., ``A large scale low-background
liquid scintillator detector: the counting test facility at Gran Sasso.''
NIM A 406 (1998) p.411-426. 

\bibitem{Wright}Wright A.G., ``Determination of the multiplier gain
of a photomultiplier.'', J.Phys. E: Sci.Instrum., Vol.14, 1981, p.851-855.

\bibitem{Filters}R. Dossi, A. Ianni, G. Ranucci, O. Ju. Smirnov.
``Methods for precise photoelectron counting with photomultipliers.''
 NIM A451 (2000) 623-637. 

\bibitem{Chirikov-Zorin}I.Chirikov-Zorin, I.Fedorko, A.Menzione,
M.Pikna, I.Sykora, S.Tokar. ''Method for precise analysis of the
metal package photomultiplier single photoelectron spectra'', NIM
A456 (2001) 310-324. 

\bibitem{5}G.Bacchiocchi , A. Brigatti, R. Dossi, A. Ianni, O. Smirnov.
``The Earth's magnetic field compensation in the Borexino Phototubes
facility.'' LNGS preprint INFN/TC-97/35, 1997. Available at http://lngs.infn.it/.

\bibitem{Cat}Thorn EMI Electron Tubes catalogue,1993. 

\bibitem{Database}O.Smirnov. CTF-II Photomultipliers database. BOREXINO
PMT working group home page at http://www.pcbx01.lngs.infn.it
\end{thebibliography}
\end{document}